\documentclass[%
reprint,
superscriptaddress,
footinbib,
amsmath,amssymb,
aps,pre,
nobibnotes
]{revtex4-2}

\usepackage[english]{babel}
\usepackage[utf8x]{inputenc}
\usepackage[colorinlistoftodos]{todonotes}
\usepackage{enumerate}
\usepackage{graphicx}% Include figure files
\usepackage{dcolumn}% Align table columns on decimal point
\usepackage{bm}% bold math
\usepackage{textcomp}
\usepackage{xcolor}

\usepackage[pagewise,columnwise]{lineno}
\newcommand{\pT}{p_\text{T}}
\newcommand{\W}{\mathcal{W}}

\begin{document}

\preprint{APS/123-QED}

\title{Internal energy of the medium formed in pp collisions}


\author{J. Alonso Tlali}
\affiliation{Facultad de Ciencias F\'isico Matem\'aticas, Benem\'erita Universidad Aut\'onoma de Puebla, Apartado Postal 165, 72000 Puebla, Pue., M\'exico}

\author{D. Rosales Herrera}
\affiliation{Facultad de Ciencias F\'isico Matem\'aticas, Benem\'erita Universidad Aut\'onoma de Puebla, Apartado Postal 165, 72000 Puebla, Pue., M\'exico}

\author{J. R. Alvarado García}
\email{j.ricardo.alvarado@cern.ch}
\affiliation{Facultad de Ciencias F\'isico Matem\'aticas, Benem\'erita Universidad Aut\'onoma de Puebla, Apartado Postal 165, 72000 Puebla, Pue., M\'exico}

\author{A. Fern\'andez T\'ellez}
\affiliation{Facultad de Ciencias F\'isico Matem\'aticas, Benem\'erita Universidad Aut\'onoma de Puebla, Apartado Postal 165, 72000 Puebla, Pue., M\'exico}

\author{C. Pajares}
\affiliation{Departamento de Física de Partículas and Instituto Galego de Física de Altas Enerxías, Universidad de Santiago de Compostela, Santiago de Compostela, E-15782, Spain}

\author{J. E. Ram\'irez}
\email{jhony.eredi.ramirez.cancino@cern.ch}
\affiliation{Facultad de Ciencias F\'isico Matem\'aticas, Benem\'erita Universidad Aut\'onoma de Puebla, Apartado Postal 165, 72000 Puebla, Pue., M\'exico}


\begin{abstract}
In this paper, we perform a thermostatistical analysis of the charged particles produced in minimum bias pp collisions across different energies available at the LHC-CERN reported by the ALICE Collaboration. To this end, we adopted the normalized $\pT$ spectrum as the probability density function governing the $\pT$ microstates of the produced hadrons, from which we can compute the entropy following the Shannon definition or the average $\pT$ as the Bjorken-type energy estimator. 
Additionally, we estimate the temperature from the slope of the $\pT$ spectrum at low $\pT$ values.
We use the Hagedorn and Tricomi functions to describe the $\pT$ spectrum within a nonextensive approach
and determine consistent thermodynamic frameworks by solving the fundamental relation $dE/dT=C$, where $C=TdS/dT$ is the heat capacity.
We found that the internal energy behaves nonlinearly at LHC energies, indicating that the count of degrees of freedom is more complex than simply adding the number of flavors present in the hadron production.
The main implication of our results is that the systems produced in pp collisions depart from the ideal gas picture and explain why the production of high $\pT$ hadrons has always been observed, even in earlier experiments, which may also encompass other experiments colliding e$^-$p or e$^+$e$^-$.
\end{abstract}

\maketitle

\section{Introduction}\label{sec:intro}

The study and determination of quark matter properties is a hot topic in the high energy physics community, particularly in understanding the thermodynamics of these systems \cite{Tsallis:1998ws,Rischke:2003mt,Wilk:2014sza,alba2014freeze,Tripathy:2016hlg,Bhattacharyya:2021rcm,ALICE:2022wpn}. %\cite{Waqas:2025ihp}
A fundamental approach for describing the strongly interacting matter involves establishing an equation of state (EOS) that can characterize such systems across variations of the thermodynamic variables \cite{VOGT2007221}. 
This description relies on fundamental relations among the local temperature $T$, energy density $\varepsilon$, pressure $\mathcal{P}$, and other physical quantities, from which additional variables can be derived through standard methods \cite{Kapusta:2006pm}.
Within this context, the Landau thermodynamic framework offers a method for deriving an EOS for nuclear matter across a broad range of conditions: from ultrarelativistic particle collisions to the high baryon density regime characteristic of neutron stars and other compact objects \cite{Landau:1980mil,Baym:2017whm}.
These scenarios present distinct features that must be incorporated into the EOS.
In particular, the finite value of the baryon chemical potential ($\mu_B$) marks a substantial difference, requiring the EOS to incorporate the baryon number density $n_B$ \cite{Fukushima:2010bq,Oertel:2016bki}. 
In contrast, for ultrarelativistic matter at vanishing baryon density, the fundamental relation $\varepsilon = 3\mathcal{P}$ serves as an ideal baseline for the thermodynamics. 
Moreover, deviations from the ideal ultrarelativistic limit can be incorporated by studying the trace anomaly, which quantifies the breaking of scale symmetry in QCD \cite{Collins:1976yq,Ji:1994av}.%la segunda referencia habla detalladamente de la anomalía de la traza pero no de la EoS.

One of the most successful techniques for studying QCD matter is lattice QCD 
\cite{Wilson:1974sk}.
This approach provides a reliable EOS that matches smoothly to hadron resonance gas models at low temperatures through the analysis of trace anomaly behavior \cite{Philipsen:2012nu}.  
At vanishing baryon chemical potential, the energy density and other rheological properties show a rapid increase in the high-temperature regime \cite{Borsanyi:2010cj,HotQCD:2014kol}.
Specifically, the behavior of $\varepsilon$ aligns with the Stefan-Boltzmann relation  $\varepsilon = \sigma_{SB} T^4$, where $\sigma_{SB}$ depends on the number of active degrees of freedom $\sigma_{SB} = \pi^2 N/30 $.
For instance, in the case of a pion gas, $N=3$, which are considered as bosonic degrees of freedom \cite{Hagedorn:1983wk}. 
For the quark gluon plasma phase $N = g_B + \frac{7}{8}g_F$, where $g_B = 16$ (accounting for 2 helicity states and 8 color combinations for gluons), and $g_F = 12 N_f$ (accounting for quarks and antiquarks with 3 colors and 2 spin states each, summed over the number of quark flavors $N_f$) \cite{Hagedorn:1983wk}.

Complementary to lattice calculations, functional approaches provide direct non-perturbative calculations at finite temperature and chemical potential, focusing on the finite-density part of the QCD EOS \cite{Lu:2023msn}.
Such techniques include the study of the Dyson-Schwinger equations and renormalization group methods \cite{Gao:2020fbl,Gunkel:2021oya,Fu:2019hdw,Lu:2023msn}.

Other theoretical approaches establish an EOS across different temperature and density regimes \cite{Rischke:2003mt}. 
For instance, hard-thermal-loop perturbation theory addresses high-temperature behavior \cite{Andersen:2004fp}, while quasiparticle models incorporate medium effects through temperature-dependent self-energies \cite{Peshier:1999ww,Bluhm:2004xn}.
Polyakov-enhanced chiral models couple chiral and deconfinement order parameters \cite{Fukushima:2003fw,Roessner:2006xn}, and holographic approaches provide nonperturbative constraints from string theory \cite{Policastro:2001yc,Buchel:2007mf,Casalderrey-Solana:2011dxg}.
More recently, Bayesian analyses have systematically extracted transport coefficients by comparing multi-stage hydrodynamical models with heavy-ion collision data, advancing in the quantitative determination of properties concerning the QCD matter within the thermodynamic framework \cite{Bernhard:2016tnd,JETSCAPE:2017eso,JETSCAPE:2020mzn}.

From the experimental perspective, the energy density in heavy-ion collisions is commonly estimated using the Bjorken formula $\varepsilon = \langle m_\text{T} \rangle  \mu / \tau A$, which relates the transverse particles' energy flux $\langle m_T \rangle \mu  = \langle m_T \rangle  dN/d\eta$ per unit transverse area $A$ evolving according to its proper time $\tau$ \cite{Bjorken:1982qr}.
%While the transverse area is experimentally accessible through the Glauber model for heavy-ion collisions 
In heavy-ion collisions, the transverse area is dominated by the nuclear geometry, while the internal structure of the colliding nuclei does not contribute significantly~\cite{Bialas:1976ed,Loizides:2016djv}. %Loizides:2016djv
In these systems, the Glauber model provides a centrality dependent estimation of $A$ over a wide range of collision events~\cite{Miller:2007ri}.
In contrast, for small collision systems, no comparable nuclear length scale is available, and the interaction area is inferred from assumptions about the partonic structure of the system~\cite{Mantysaari:2020axf}.
For these cases, several studies rely on estimating the transverse interaction area in small systems as a function of charged-particle multiplicity.
For instance, the interaction radius can be estimated from the gluon saturation scale within the Color Glass Condensate framework, describing the multiplicity dependence of the average transverse momentum in high multiplicity pp and pA collisions~\cite{McLerran:2013oju}.
Similarly, the Color String Percolation Model relates the transverse area to the color string density through a temperature scaling of the multiplicity using the IP-Glasma model for pp collisions and the Glauber model for nucleus-nucleus collisions~\cite{Mishra:2022kre}.
In both cases, the resulting transverse area remains dependent on the model describing the partonic structure. This limitation becomes more relevant in minimum bias pp collisions because the interaction area is not determined on an event-by-event basis.

Consequently, constructing an EOS for small systems represents a limitation in extending the Landau framework, particularly regarding volume scaling when matching to experimental data.
Additionally, small collision systems exhibit significant finite-size effects incompatible with the thermodynamic limit assumed for large systems, where fluctuations average out and extensive variables scale with system size \cite{Nagle:2018nvi,Garcia:2022ozz}. 
Therefore, developing a fundamental EOS that accurately describes experimental particle collision data across the range from small to large systems requires alternative theoretical considerations beyond the established approaches.

%propuesta JE del párrafo anterior
An alternative method for developing a consistent thermostatistics for the medium created in ultrarelativistic collisions involves adopting the general ensemble theory, which treats the $\pT$ spectrum of the produced particles as a probability density function \cite{Herrera:2024zjy,Lopez:2025iki}.
Here, the microstates of the system correspond to transverse momentum states instead of the usual energy states.
In this way, the computation of the Shannon entropy can provide thermodynamic information about the system where the particles are created.
For instance, recent results show that the entropy of minimum bias pp collisions increases with the center of mass energy \cite{Herrera:2024zjy}, indicating an enhancement of high $\pT$ particles, but also an increment of the microscopic degrees of freedom, similar to classical thermodynamics.
In the experimental high energy physics community, the slope of the $\pT$ spectrum at low $\pT$ values on a log-log plot is considered a temperature-like parameter \cite{ALICE:2022wpn}. 
This interpretation stems from the exponential behavior of the $\pT$ spectrum observed at low $\pT$, similar to the Boltzmann distribution, where the inverse of the exponential decay constant is used as the temperature \cite{DiasdeDeus:2006xk}.
One major issue with this definition arises from selecting the appropriate range for performing the fit to the data, where the temperature is extracted.
Better estimations can be achieved by using fitting functions that accurately describe the entire $\pT$ spectrum, where the temperature can be determined by analyzing the asymptotic behavior at low $\pT$ values \cite{Pajares:2022uts,Garcia:2023eqg}.

At this point, we have introduced the entropy and temperature as two thermodynamic observables for describing the particle production, which are related through the fundamental relation $\partial E/ \partial S=T$, with $E$ being the internal energy. 
In this way, the internal energy cannot be an arbitrary function; it must be determined according to the pre-established definitions of entropy and temperature.

%aim and scope JE
In this paper, we aim to determine the internal energy $E$ as a function of the temperature within the nonextensive description of the charged particles produced in pp collisions at minimum bias and midrapidity conditions using the string fragmentation framework provided by the Hagedorn and Tricomi models. 
This coarse-graining approach has proven to be a fundamental requirement for describing high $\pT$ particle production in ultrarelativistic collisions, modeling the hard QCD processes as rare events arising from large fluctuations in the strength of the color interactions \cite{RosalesHerrera:2026fmg}.
The physical motivation for including such events comes from hard partonic scattering, heavy quark production, and hard gluon emission processes, similarly to those implemented in the Lund model \cite{Andersson:1997xwk}.
To contrast the thermodynamics derived from the slope of the $\pT$ spectrum and the Shannon entropy, we also estimate a thermodynamically consistent entropy by choosing the average $\pT$ as the internal energy per particle and retaining the temperature definition.

The rest of the paper is organized as follows.
In Sec.~\ref{sec:fragmentation}, we discuss in detail the particle production mechanism usually adopted for color string fragmentation models. Here, the $\pT$ spectrum is derived from the convolution of the Schwinger mechanism and heavy tailed distributions for the string tension fluctuations.
Additionally, we commented on the statistics of the $\pT$ spectrum and the temperature fluctuations obtained from the superstatistic approach.
Section~\ref{sec:data_analysis} discusses the methodology for analyzing the reported data. In particular, we focus on analyzing the data on the production of charged particles in minimum bias pp collisions at the LHC reported by the ALICE Collaboration.
We devoted Sec.~\ref{sec:results} to derive the thermostatistics of the medium formed in pp collisions by considering only the information provided by the $\pT$ spectrum.
Finally, we wrap up this work with our concluding remarks in Sec.~\ref{sec:conclusions}.

\section{Particle production in the string fragmentation model}\label{sec:fragmentation}

One way to describe the charged particle production is the color string fragmentation model, which considers the production of new neutral color pairs via the quantum tunneling of two relativistic particles interacting through an attractive linear potential, resembling a \textit{Chinese finger trap}, where the interaction becomes stronger as the particles separate \cite{Andersson:1983ia}.
Here, the magnitude of the force produced by such linear potential is interpreted as the magnitude of the color interaction, usually named the tension of the color string.
Under these considerations, the probability of observing the creation of a new pair is 
\begin{equation}
    \frac{dN}{d\pT^2} \propto e^{-\pi \pT^2/x^2},
    \label{eq:SM}
\end{equation}
where $\pT$ is the transverse momentum of the hadron produced and $x^2$ is the tension of the string \cite{Andersson:1983ia,Andersson:1997xwk,RosalesHerrera:2026fmg}.
Equation~\eqref{eq:SM} is known as the Schwinger mechanism since it resembles the derivation of Schwinger for the description of the electron-positron pair creation from the vacuum in the presence of a very intense electric field \cite{Schwinger:1962tp}, but also Eq.~\eqref{eq:SM} is similar to other pair creation mechanisms, such as the Hawking-Unruh effect, which describe the particle-antiparticle pairs emitted near the event horizon of a black hole \cite{Hawking:1975vcx,Unruh:1976db}. 

As we commented above, the string tension can fluctuate along the evolution of the system after the collision. Thus, the particle production must account for
the Schwinger mechanism 
along with fluctuations in the color field intensity of the interacting partons (string tension) \cite{Schwinger:1962tp, Bialas:1999zg}.
Therefore, the $\pT$ spectrum is computed as follows:
\begin{equation}
    \frac{dN}{d\pT^2} \propto \int_0^\infty e^{-\pi\pT^2/x^2} P(x) dx=\W(\pT), \label{eq:conv}
\end{equation}
where the Gaussian kernel corresponds to the Schwinger mechanism and $P(x)$ is the probability density function describing the tension fluctuations.
The result of the convolution in Eq.~\eqref{eq:conv} has been denoted as $\W$ to improve notation.
In particular, the following string tension fluctuations
\begin{align}
    P_\text{Hag}(x)&= \frac{m p_0^m \pi^{\frac{m-1}{2}}  }{x^{m+1} }
    U\left(\frac{m+1}{2},\frac{1}{2},\frac{\pi p_0^2}{x^2}\right) \label{eq:PHag} ,\\
    P_\text{U}(x)&= \mathcal{N}\left( 1+\frac{(q-1)x^2}{2\sigma^2}  \right)^\frac{1}{1-q} \label{eq:PU},
\end{align}
lead to 
\begin{align}
   \W_\text{Hag}(\pT) &= \left(1+\frac{\pT}{p_0} \right)^{-m} , \label{eq:WHag}\\
    \W_U(\pT)&= U\left( \frac{1}{q-1}-\frac{1}{2}, \frac{1}{2}, \frac{\pi (q-1)\pT^2}{2\sigma^2}  \right) \label{eq:WU},
\end{align}
which correspond to the Hagedorn and Tricomi functions, respectively \cite{Herrera:2024tyq,RosalesHerrera:2026fmg}, where the $U$ function is defined as \cite{arfken}
\begin{equation}
    U(a, b, z)=  \frac{1}{\Gamma(a)}\int_0^\infty e^{-zt} t^{a-1}(1+t)^{b-a-1} dt.\label{eq:U}
\end{equation}
The free parameters $p_0$, $m$, $q$, and $\sigma$ should be determined by fitting the functions $\W_\text{Hag}$ and $\W_U$ to the experimental data.

Note that $P_U$ in Eq.~\eqref{eq:PU} is the well-known $q$-Gaussian distribution, frequently used to describe complex phenomena in nature \cite{Tsallis:1987eu,Tsallis:2009zex}.
In this context, the $q$ parameter captures the enhancement of high string tension values due to processes that can be interpreted as intense color interactions, such as gluon fusion, hard scattering, vector meson production, heavy quark production, among others, that can no longer be described by a Gaussian distribution~\cite{Pajares:2022uts}.
Thus, the $q$-Gaussian distribution incorporates these processes into the color string fragmentation model within the stochastic picture of QCD~\cite{Pajares:2022uts}.
It was discussed that $1<q<3/2$ to guarantee the convergence of $\langle\pT\rangle$ \cite{budini}. It is not expected to observe $q<1$, because it implies that the string tensions would be constrained to a finite interval $x^2 \leq 2\sigma^2/(1-q)$, limiting the maximum $\pT$ value of the high $\pT$ particle production \cite{Pajares:2022uts,Garcia:2023eqg}.
We must emphasize that both $P_\text{Hag}$ and $P_U$ are heavy-tailed distributions, which are essential for accurately describing the power-law tail of the $\pT$ spectrum.
The power-law tail of the tension fluctuation is particularly important in the color string fragmentation model because it increases the probability of observing large tension values, corresponding to intense color interactions.
In particular, the Pareto-type behavior of the fluctuations' tail is required for the Koba-Nielsen-Olesen (KNO) scaling~\cite{RosalesHerrera:2026fmg}.
In this sense, our coarse-grained approach describes partonic collisions at the initial state of the system, consistent with hard partonic scattering in perturbative QCD, and incorporates the production of high-$\pT$ hadrons as rare events.
The consequence of considering Eqs.~\eqref{eq:PHag} and~\eqref{eq:PU} is that the initial state of the system becomes nonextensive at the level of the color interactions~\cite{Herrera:2024tyq}.
In the context of the Tsallis statistics, the nonextensivity emerges when $q$ takes values differing from one and marks the onset of the nonadditivity of the entropy for systems satisfying $q$-exponential and $q$-Gaussian distributions. This property is particularly consistent with the nonadditive character of the vectorial color sum rules~\cite{Braun:2015eoa,Bautista:2019mts,Garcia:2023eqg}.
The main implication of this result is the existence of temperature fluctuations in the medium where the particles are produced, and then the final state admits an out of equilibrium picture~\cite{Herrera:2024tyq}.

It is worth noting that the Hagedorn function reproduces the $q$-exponential Tsallis distribution through the following change of variables:
$m=1/(q_e-1)$ and $p_0=\lambda/(q_e-1)$ \cite{Wilk:1999dr,Herrera:2024tyq}. %ellos también hacen la identificación
Here, we have used the subscript $e$ to denote a different $q$ parameter in the $q$-exponential Tsallis function than in the $q$-Gaussian distribution.
Therefore, the Hagedorn, $q$-exponential Tsallis, and Tricomi functions arise from a nonextensive approach to the string tension fluctuations, corresponding to a nonthermal description of the initial state of the collision system \cite{Herrera:2024tyq}.

\subsection{Temperature and temperature fluctuations}

Interestingly, the asymptotic behavior at low $\pT$ of $\W_\text{Hag}$ and $\W_U$ is an exponential decay in the form $e^{-\pT/T}$, from which we define the temperature as
\begin{align}
    T_\text{Hag}&=  \frac{p_0}{m} ,\label{eq:THag}\\
    T_U &= \sigma \frac{\Gamma\left( \frac{1}{q-1}-\frac{1}{2} \right)}{\sqrt{2\pi (q-1)}\Gamma\left( \frac{1}{q-1} \right)}, \label{eq:TU}
\end{align}
for the Hagedorn and Tricomi functions, respectively \cite{Herrera:2024zjy}.
%In the rest of this paper, we adopt Eqs.~\eqref{eq:THag} and~\eqref{eq:TU} as the temperature for the model corresponding to $\W_\text{Hag}$ and $\W_U$, respectively.
Note that these temperature definitions also incorporate information about the nonextensivity of the system through the parameters $m$ and $q$. 
In this way, the temperature depends not only on the details of the soft part of the $\pT$ spectrum but also includes the information of the hard part, which is inherently present in the experimental $\pT$-spectrum.
Additionally, the effective temperature %defined as the slope of the spectrum at low $\pT$ values, 
characterizes the kinetic freeze-out stage.
This definition differs from other statistical-thermal models where the chemical freeze-out temperature is extracted from the hadron abundances~\cite{Becattini:2005xt}.

On the other hand, the superstatistics framework describes nonequilibrium systems with complex dynamics in stationary states, characterized by fluctuations of an intensive quantity across the system~\cite{Wilk:1999dr,Beck:2002lef,Beck:2003kz}.
By considering this approach, we can infer fundamental thermodynamic information of the nonextensive models describing the $\pT$ spectrum given by Eqs.~\eqref{eq:WHag} and \eqref{eq:WU}.
In this framework, the main idea is to express the particle production as a convolution between the thermal distribution ($e^{-\pT/T}$) and temperature fluctuations ($\mathcal{T}(T)$), that is
\begin{equation}
    \frac{dN}{d\pT^2} \propto \int_{0}^\infty e^{-\pT/T}\mathcal{T}(T) dT.
\end{equation}
In particular, the following probability density functions
\begin{align}
  \mathcal{T}_\text{Hag}(T) &= \frac{\Gamma(1/T; m, p_0)}{T^2},\\
   \mathcal{T}_U(T) &=
  \frac{2}{T^3}\, \Gamma\left( \frac{1}{T^2};
 \frac{1}{q-1} - \frac{1}{2}, \frac{\sigma^2}{2 \pi (q-1)}    \right),
 \label{eq:TU_fluct}
\end{align}
reproduce the Hagedorn~\eqref{eq:WHag} and Tricomi~\eqref{eq:WU} models, respectively \cite{Wilk:1999dr,Herrera:2024tyq}.
The main implication of the temperature fluctuations is that the system now shows local temperature variations in small regions along it, indicating that the system is out of thermal equilibrium. In contrast, the thermal distribution is obtained by considering the Dirac delta function as temperature fluctuations, which implies that the system has a uniform temperature throughout.

\subsection{$\pT$ spectrum statistics}

One advantage of having the functions \eqref{eq:WHag} and \eqref{eq:WU} corresponding to the $\pT$ spectrum of the Hagedorn and Tricomi models is the possibility to derive closed formulas for some observables, such as the statistics. 
In particular, the $\ell$-moment of the $\pT$ spectrum is computed in the standard way:
\begin{equation}
\langle \mathcal{P}_T^\ell \rangle= \frac{ \int_0^{\infty}  p_T^\ell  \W(\pT) dp_T}{\int_0^{\infty} \W(\pT) dp_T}.
\label{eq:ptn}
\end{equation}
Here, we must distinguish the moments of the $\pT$-spectrum from the statistics reported by the high energy community, which involves the differential contributions from the phase space factor \cite{Hagedorn:1983wk,Bylinkin:2014fta}.
Thus,
\begin{equation}
    \langle \pT^\ell \rangle=\frac{\int \pT^\ell \W(\pT) 2\pi \pT d\pT}{\int \W(\pT) 2\pi \pT d\pT} =\frac{\langle \mathcal{P}_\text{T}^{\ell+1} \rangle}{\langle\mathcal{P}_\text{T}\rangle}. \label{eq:momentslab}
\end{equation}
The latter definition is also equivalent to considering $\pT^2$ as the random variable.
In particular, the average $\langle \pT \rangle$ is
\begin{subequations}
\begin{align}
    \langle \pT \rangle_\text{th} &= 2T_\text{th},\\
     \langle \pT \rangle_\text{Hag} &=\frac{2m}{m-3} T_\text{Hag}, \label{eq:avgpTHag}\\
     \langle \pT \rangle_\text{U}&= \frac{(q-1)(3q-5)}{(2-q)(2q-3)} \left( \frac{\Gamma\left( \frac{1}{q-1}\right) }{\Gamma\left( \frac{1}{q-1} - \frac{1}{2}\right)} \right)^2 T_U \label{eq:avgpTU},
\end{align}
\end{subequations}
for the thermal, Hagedorn, and Tricomi models.
Note that in all cases, the average $\pT$ is proportional to the temperature in the corresponding model. However, for the Hagedorn and Tricomi models, the proportionality factors only depend on the parameter characterizing the deviations of the systems from the thermal picture.
In any case, this proportionality factor is not constant for minimum bias pp collisions.

\section{Data analysis}\label{sec:data_analysis}

The data sets we analyzed correspond to the production of charged particles in pp collisions under minimum bias conditions at various center of mass energies reported by the ALICE Collaboration \cite{ALICE:2010syw, ALICE:2022xip}.
For these data sets, the kinematical cuts are: $|\eta|<0.8$ and $0.15\text{ GeV}<\pT<10\text{ GeV}$.
Under these conditions, the baryon chemical potential is zero at LHC energies \cite{ALICE:2018pal}.
Therefore, $\mu_B$ does not contribute to the internal energy computation.
Additionally, it has been demonstrated that minimum bias pp collisions contain a negligible contribution from particles produced by collective phenomena~\cite{RosalesHerrera:2025zqt}. 
In such cases, we can consider that the production of charged particles results from the fragmentation of color strings.

\begin{figure*}[ht]
    \centering
\includegraphics{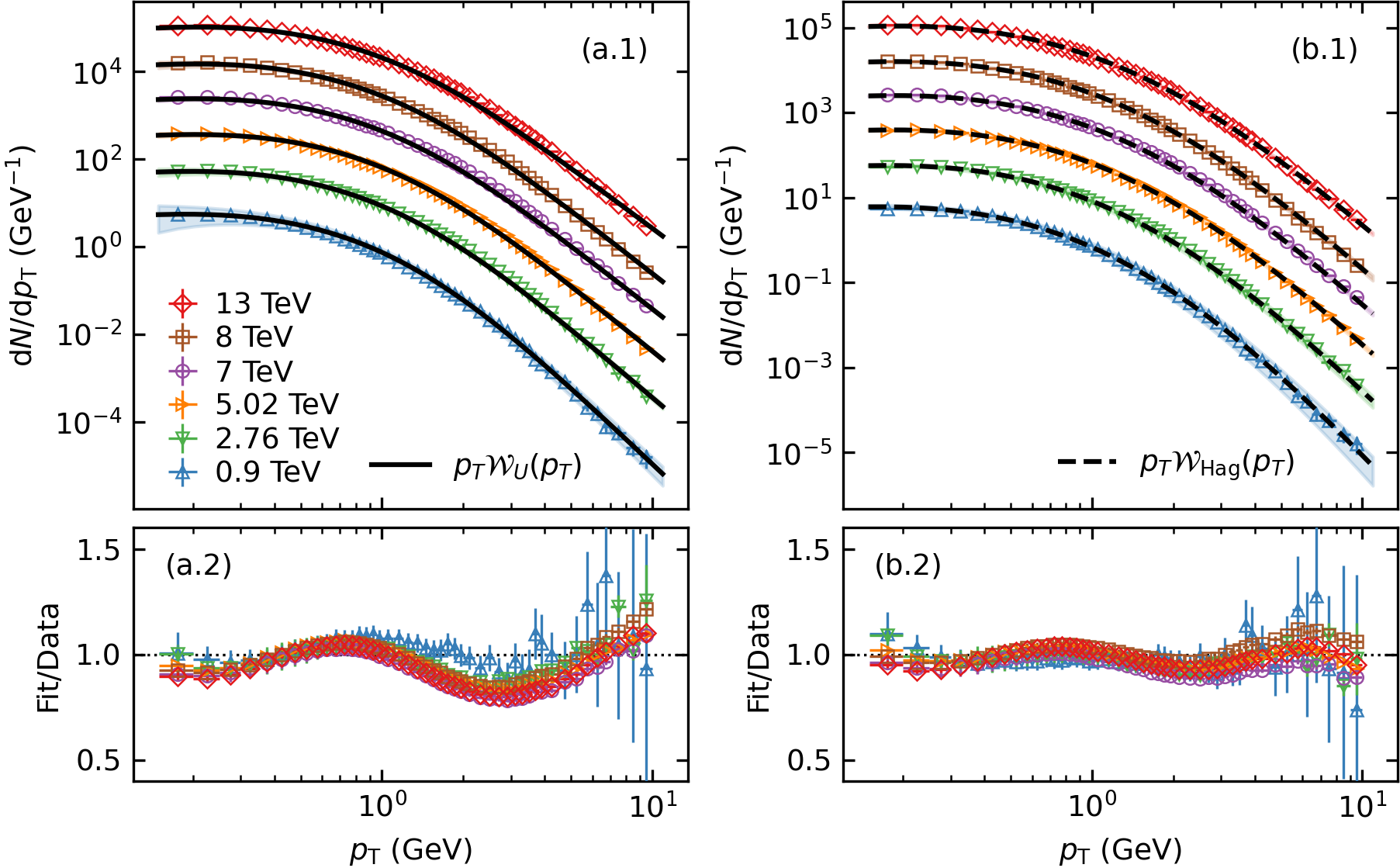}
    \caption{Fits (lines) of (a.1) $\pT\W_U$ and (b.1) $\pT\W_\text{Hag}$ to experimental data (figures) of the $\pT$ spectrum for charged particle production in minimum bias pp collisions at different center of mass energies. 
    The shaded regions correspond to the 95\% uncertainty propagation of fits.
    The corresponding fit to data ratios are shown in lower panels for the (a.2) $U$ and (b.2) Hagedorn function.}
    \label{fig:fits}
\end{figure*}

The fits to the data were carried out by using the software ROOT 6, considering the following procedure.
In the first place, note that the data reported by the ALICE Collaboration correspond to $dN/d\pT$ instead of $dN/d\pT^2$. 
Therefore, the fitting functions should consider an extra $\pT$ factor to be $\pT \W_\text{Hag}$ and $\pT \W_\text{U}$ instead of \eqref{eq:WHag} and \eqref{eq:WU}, respectively.
The fitting procedure excluded an intermediate region typically from $1$ to $6$ GeV where collective effects become relevant~\cite{RosalesHerrera:2025zqt}.
The main contribution to the particle production in this regime comes from the production of softened hadrons, which can not be described by the string fragmentation model~\cite{RosalesHerrera:2025zqt}. 
However, it is shown that the production of softened hadrons vanishes as the average multiplicity decreases~\cite{RosalesHerrera:2025zqt}. 
Therefore, the production of softened hadrons is expected to be negligible at the lowest center of mass energies. 
This is the main reason for excluding these effects from the analysis.
%This region is excluded because it is affected by collective effects and their description requires incorporating a softened hadron production model, which is beyond the string fragmentation mechanism~\cite{osalesHerrera:2025zqt}}
The excluded interval is determined by minimizing the $\chi^2$/NDF considering only the low and high $\pT$ regions of the $\pT$ spectrum.
%The goodness of fit was assessed using the Kolmogorov-Smirnov test (see Appendix \ref{app:KS}), which indicates excellent agreement with $p > 0.90$ for all datasets and both models.
In Fig.~\ref{fig:fits}, we show the fits to experimental data for the $U$ and Hagedorn functions with their corresponding fit to data ratios.

We performed Kolmogorov-Smirnov (KS) tests to assess the quality of our fits. This is done by comparing the cumulative distribution functions (CDFs) of the fitted models with the empirical CDF from ALICE data. 
The KS statistic $D_{\rm KS}$ is defined as the maximum absolute deviation between the empirical and theoretical CDFs.

The empirical CDF is computed from the normalized $\pT$ spectrum over the interval $[0.15, 10]$ GeV using $N = 47$ bin edges. 
For the Hagedorn function, the CDF admits a closed-form expression that depends on the integration limits. 
For the Tricomi function, the CDF is computed numerically using the GSL library.
The maximum deviation $D_{\rm KS}$ is evaluated at the bin centers of the experimental $\pT$ spectrum, and the corresponding $p$-value is calculated using the asymptotic Kolmogorov distribution with continuity correction via the ksone function imported from scipy.stats in Python. 
\begin{figure*}
    \centering
    \includegraphics{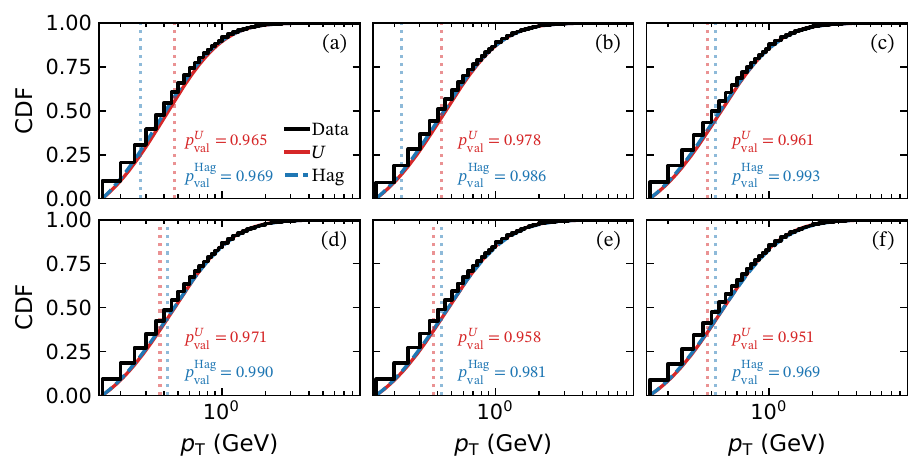}
    \caption{Cumulative distribution functions (CDFs) for the experimental data (solid step lines), Tricomi function $\mathcal{W}_U$ (solid lines), and Hagedorn function $\mathcal{W}_{\rm Hag}$ (dashed lines) at the center of mass energies of (a) 0.9, (b) 2.76, (c) 5.02, (d) 7, (e) 8, and (f) 13 TeV. 
    The vertical dotted lines indicate the $\pT$ bin where the $D_{\rm KS}$ has its maximum for each model. 
    The corresponding $p$-values are shown for both functions.}
    \label{fig:KS}
\end{figure*}
Figure~\ref{fig:KS} shows the experimental and theoretical CDFs along with the corresponding $p$-values for all energy datasets. 
In all cases, $D_{\rm KS} < 0.021$, below the critical value $D_{\rm crit}(5\%) = 0.198$ at the 95\% confidence level \cite{Massey1951}.
In consequence, in all cases the $p$-values exceed 0.90, meaning that we cannot reject the null hypothesis and the fitted functions are consistent with the data.

\begin{figure}
    \centering
\includegraphics{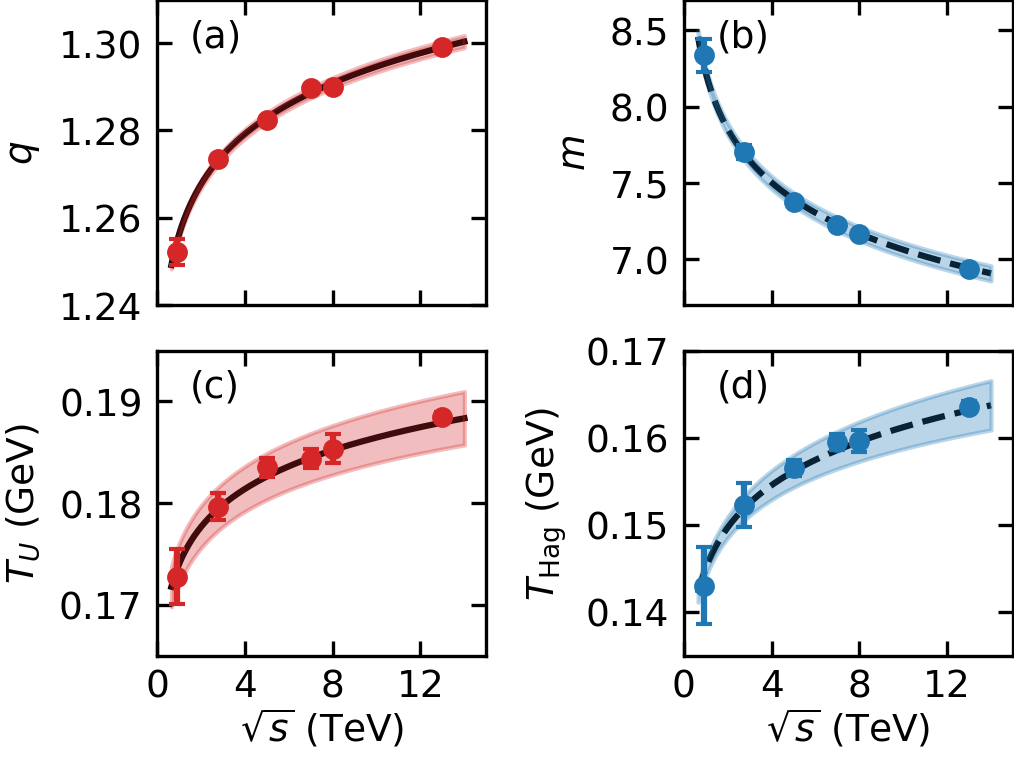}
    \caption{Behavior of (a) $q$, (b) $m$, (c) $T_U$, and (d) $T_\text{Hag}$ as a function of the center of mass energy. The lines correspond to the power law parametrization for the $U$ (solid lines) and Hagedorn function (dashed lines) with their corresponding uncertainty propagation (shaded region).}
    \label{fig:params}
\end{figure}

In Fig.~\ref{fig:params}, we plot our results of the fitting parameters $m$ and $q$ together with the estimation of the temperature for each model.
Here, it is important to note that the free parameters and temperature are not independent by definition (see Eqs.~\eqref{eq:THag} and \eqref{eq:TU}). From now on, we will disregard $p_0$ and $\sigma$ in favor of the temperatures $T_\text{Hag}$ and $T_U$ because the thermodynamic analysis requires derivations and integrations that explicitly depend on the temperature.
In all cases, we observe that the free parameters follow a power law trend as functions of the center of mass energy given by
\begin{equation}
    X=a_X \left( \frac{\sqrt{s}-\sqrt{s_\text{min}}}{\sqrt{s_0}} \right)^{c_X}, \label{eq:parametrization}
\end{equation}
with $\sqrt{s_0}=1$ TeV and $\sqrt{s_\text{min}}=2m_p$ is the minimal center of mass energy for two colliding protons. Equation \eqref{eq:parametrization} allows us to write down the nonextensivity parameters $m$ and $q$ as functions of $T_\text{Hag}$ and $T_U$, respectively.
In Table~\ref{tab:fit_values}, we show the results obtained for the parametrization \eqref{eq:parametrization} of $m$, $q$, $T_\text{Hag}$, and $T_U$.

%Fitted values of the power law describing the $U$ and Hagedorn parameters as a function of $\sqrt{s}$.

\begin{table}
\caption{Fitted values of the power law dependence on $\sqrt{s}$ of the parameters corresponding to the $U$ and Hagedorn functions.
}
%    \begin{ruledtabular}
\centering
\begin{tabular}{c c c}\label{tab:fit_values}

Parameter & $a_X$ & $c_X$ \\ 
\hline
$q$ &  1.2562(9)  &  0.0131(3)   \\
$m$ &  8.22(4) &  -0.066(2)   \\
$T_U$ & 0.174(2) GeV & 0.030(5)  \\
$T_\text{Hag}$ &  0.145(1) GeV  &   0.046(6) 
    \end{tabular}%

%    \end{ruledtabular}
\end{table}

\section{Results}\label{sec:results}

It is worth noting that the functions $\W_\text{Hag}$ and $\W_U$ recapture the thermal behavior in the limit $m\to \infty$ and $q\to 1$, respectively. 
This fact suggests the existence of a specific temperature value at which $q(T_U^*)=1$ if the power law trend \eqref{eq:parametrization} stands up to the lower center of mass energies. 
This is assumed since there are no available data reported by earlier experiments consistent with the kinematical cuts reported by the ALICE Collaboration. 
%The latter indicates the existence of a thermal to nonthermal transition in the production of charged particles in pp collisions. 
Additionally, this assumption is useful for establishing the initial conditions required for estimating the internal energy.
At this point, the system recovers the ideal gas picture and then $E=T$.
We found $T_U^{*}=0.103(6)$ GeV by extrapolating the power law trend of $q$ as a function of $T_U$.
In this way, the model with the $q$-Gaussian string tension fluctuations gives a $\pT$ spectrum of the form
\begin{equation}
    \W_U=\begin{cases}
			e^{-\pT/T_U}, & \text{if $T_U\leq T_U^*$}\\
            U\left( \frac{1}{q-1}-\frac{1}{2}, \frac{1}{2}, \frac{\pi (q-1)\pT^2}{2\sigma^2}  \right), & \text{otherwise}
		 \end{cases},
\end{equation}
which also comes from the same limit for the $q$-Gaussian distribution.
In contrast, for the Hagedorn function $\W_\text{Hag}$, there is no immediate temperature $T_\text{Hag}^*$ because the parameter $m$ has no restrictions on the values it can take.
%However, we will also discuss the asymptotic behaviors and discuss a similar transition for $\W_\text{Hag}$ below.

\begin{figure*}
    \centering
\includegraphics{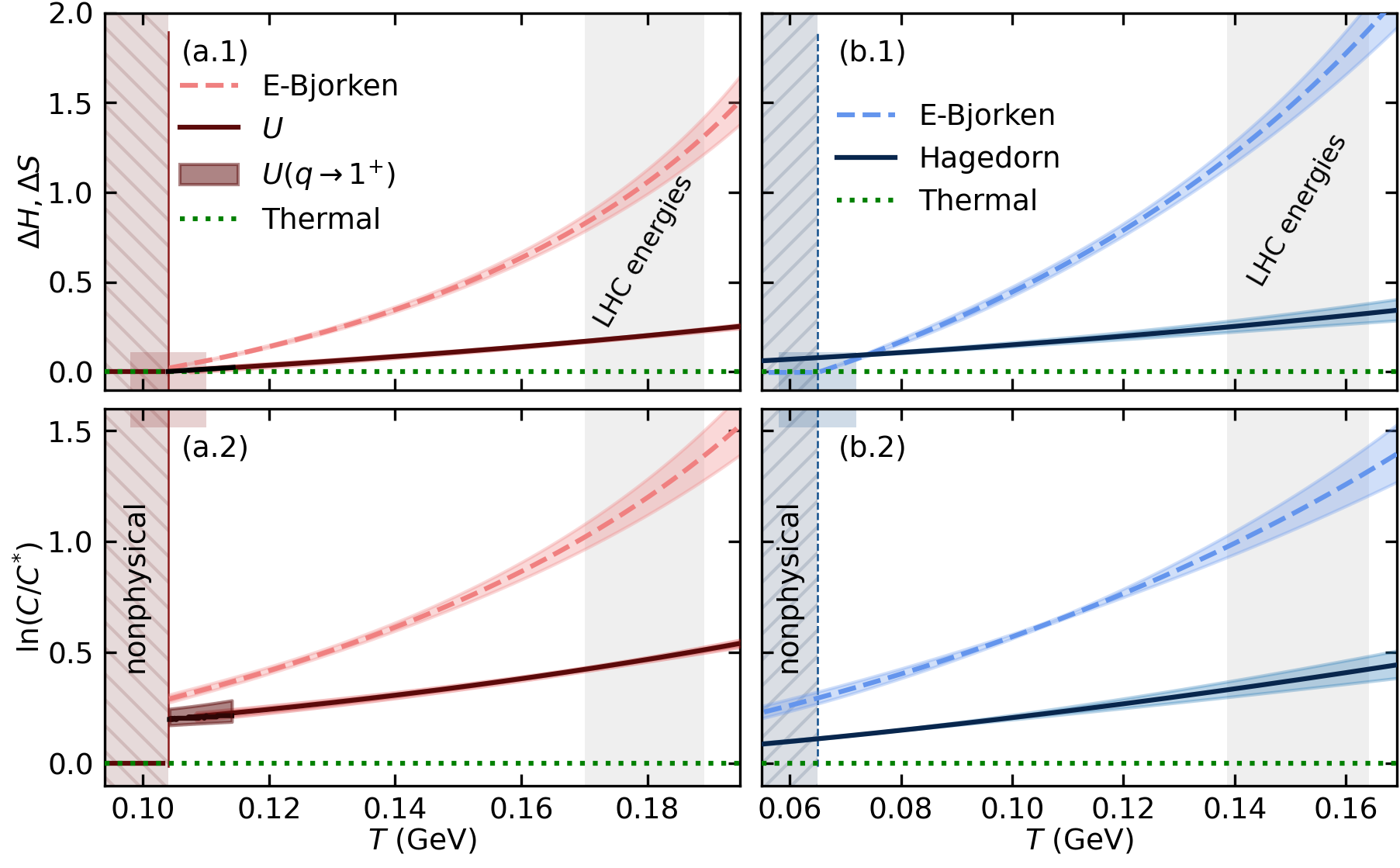}
    \caption{Temperature dependence of $\Delta H$, $\Delta S$, and heat capacity for the [(a.1) and (a.2)] $U$ and [(b.1) and (b.2)] Hagedorn functions. 
    %with their corresponding uncertainty propagation (shaded regions). 
    Solid lines correspond to the estimation using the Shannon entropy for the $U$ and Hagedorn functions.
    Dashed lines correspond to the estimation of the entropy by assuming the Bjorken-type energy for the $U$ and Hagedorn functions.
    In all cases, the shaded regions correspond to the uncertainty propagation.
    Dotted lines correspond to the estimation of the thermal model.
    The numerical calculations in the limit $q\to 1^{+}$ (bordered shaded regions) in the finite temperature interval correspond to the 95\% confidence level. 
    The vertical solid and dashed lines indicate the temperatures $T_U^*$ and $T_\text{Hag}^*$ of each model.
    The shaded gray regions illustrate the ranges of temperatures available at the LHC energies given by the $U$ and Hagedorn functions.
    The hatched regions correspond to the nonphysical region for each model.}
    \label{fig:ShannonHeat}
\end{figure*}

At this point, we have introduced the temperature as a thermodynamic variable extracted from the low-$\pT$ asymptotic behavior of the spectrum. 
However, a complete thermodynamic description requires at least two independent variables. 
To this end, we adopt the Shannon entropy as the entropy of the system \cite{shannon1948mathematical}, defined for probability density functions as follows
\begin{equation}
    H=-\int_0^\infty \hat{\W} \ln \hat{\W} d\pT,
\end{equation}
where $\hat{\W}$ denotes the normalized $\pT$ spectrum along the entire $\pT$ range. 
Note that the microstates are described by $\pT$ states rather than the conventional energy states in statistical mechanics.
In this way, we are applying the general ensemble theory to thermodynamically describe the medium where particles are produced using the data reported by the experiments.
Additionally, from thermodynamics, the heat capacity is computed as follows:
\begin{equation}
    C=T\frac{dH}{dT}.
\end{equation}
In particular, for the thermal distribution and the Hagedorn function, we can obtain the following closed forms for the entropy and heat capacity \cite{Herrera:2024zjy}: 
\begin{subequations}
\begin{align}
    H_\text{th}=& 1+\ln T_\text{th}, \label{eq:Hth} \\ 
    H_\text{Hag}=&  \frac{m}{m-1} + \ln \left( \frac{m}{m-1} \right) + \ln (T_\text{Hag}),\\ 
    C_\text{th}=&1, \label{eq:Cth}\\
    C_\text{Hag}=& 1 +  
    \frac{1-2m}{m(m-1)^2} \frac{a_m c_m}{c_{T_\text{Hag}}} \left( \frac{T_\text{Hag}}{a_{T_\text{Hag}}}  \right)^{c_m/c_{T_\text{Hag}}}.
    %\frac{1-2m}{(m-1)^2} \frac{ c_m}{c_{T_\text{Hag}}}.
\end{align}
\end{subequations}
In Eqs.~\eqref{eq:Hth} and~\eqref{eq:Cth}, the subscript th refers to the Shannon entropy and heat capacity for the thermal distribution, respectively.

On the other hand, due to the definition of the Tricomi function (see Eq.~\eqref{eq:U}), it is not possible to derive a closed formula for the entropy and heat capacity of the function $\W_U$. For this case, the entropy is numerically integrated.
Furthermore, even when $q$ is parametrized in terms of the temperature, the derivative of the entropy with respect to $T_U$ also involves calculating integrals that contain logarithms of $\W_U$ and its derivatives, which are numerically computed.
Another issue arises for $q$ values close to 1. In this case, the Tricomi function requires the computation of the gamma function for large numbers, which cannot always be performed. 
Additionally, the non-Gaussian terms in Eq.~\eqref{eq:U} rapidly diverge as $q$ approaches 1.
To address the latter, we rewrite the convolution of the Schwinger mechanism with $q$-Gaussian string tension fluctuations as the expectation of a particular function with Gaussian weights. 
This allows the convolution to be approximated numerically using Hermite quadrature, which estimates the $\pT$ spectrum for each $\pT$ bin. Consequently, the entropy is calculated using Shannon entropy over the normalized histogram.
We generate $10^6$ entropy curves as a function of $T_U$ to account for the uncertainty propagation of the power law parametrization of $q$ and $T_U$ considered as independent, normally distributed random variables. 
Subsequently, the heat capacity is estimated numerically using the five-point stencil method over each entropy curve. 
Thus, the reported entropy and heat capacity values correspond to the mode of their respective distributions at fixed $T_U$.
In Fig.~\ref{fig:ShannonHeat}, 
%we plot our results of the entropy and heat capacity for the Hagedorn and Tricomi functions.
we plot the deviation of the Shannon entropy with respect to the ideal case ($\Delta H=H-1-\ln T$) and the heat capacity for the Hagedorn and Tricomi functions.
In both cases, the entropy diverges from the thermal limit, indicating that the medium in which the charged particles are produced is nonthermal and exhibits a temperature at which the thermal limit is recovered.
%, thereby indicating a thermal-nonthermal transition.
For the estimation of  $T_\text{Hag}^*$, we extrapolate the temperature ratio $T_U/T_\text{Hag}$ until $T_U^*$, obtaining $T_\text{Hag}^*=0.63 T_U^*=0.065(7)$ GeV. 
These temperatures correspond to a center of mass energy $\sqrt{s^*} \approx 1.876$ GeV, which is computed using the proton mass value reported by the Particle Data Group \cite{ParticleDataGroup:2024cfk}.
However, this value may increase for a non-vanishing baryon chemical potential.
Note that for two colliding protons is physically impossible to access at relativistic energies below $\sqrt{s^*}$. 
Therefore, an excluded region is defined for $\sqrt{s}<\sqrt{s^*}$ (or $T<T^*$), corresponding to the nonphysical region depicted in all panels of Fig.~\ref{fig:ShannonHeat}.
%Below this threshold, there is no pp collision system, thereby defining a physically excluded region for $T<T^*$.
Moreover, all experimental pp collisions have been performed at $\sqrt{s} > \sqrt{s^*}$, satisfying the condition $T>T^*$. 
The main implication is that all the pp collision systems are in the nonthermal regime, and the production of high-$\pT$ hadrons always manifests a power-law tail in the $\pT$ spectrum.
This explains why hard processes have been consistently observed in pp collisions across all center of mass energies, even at the low ISR energies~\cite{Michael:1977hx,Banner:1973nu,British-Scandinavian:1975kbt,Hagedorn:1983wk}.

\begin{figure}
    \centering
\includegraphics{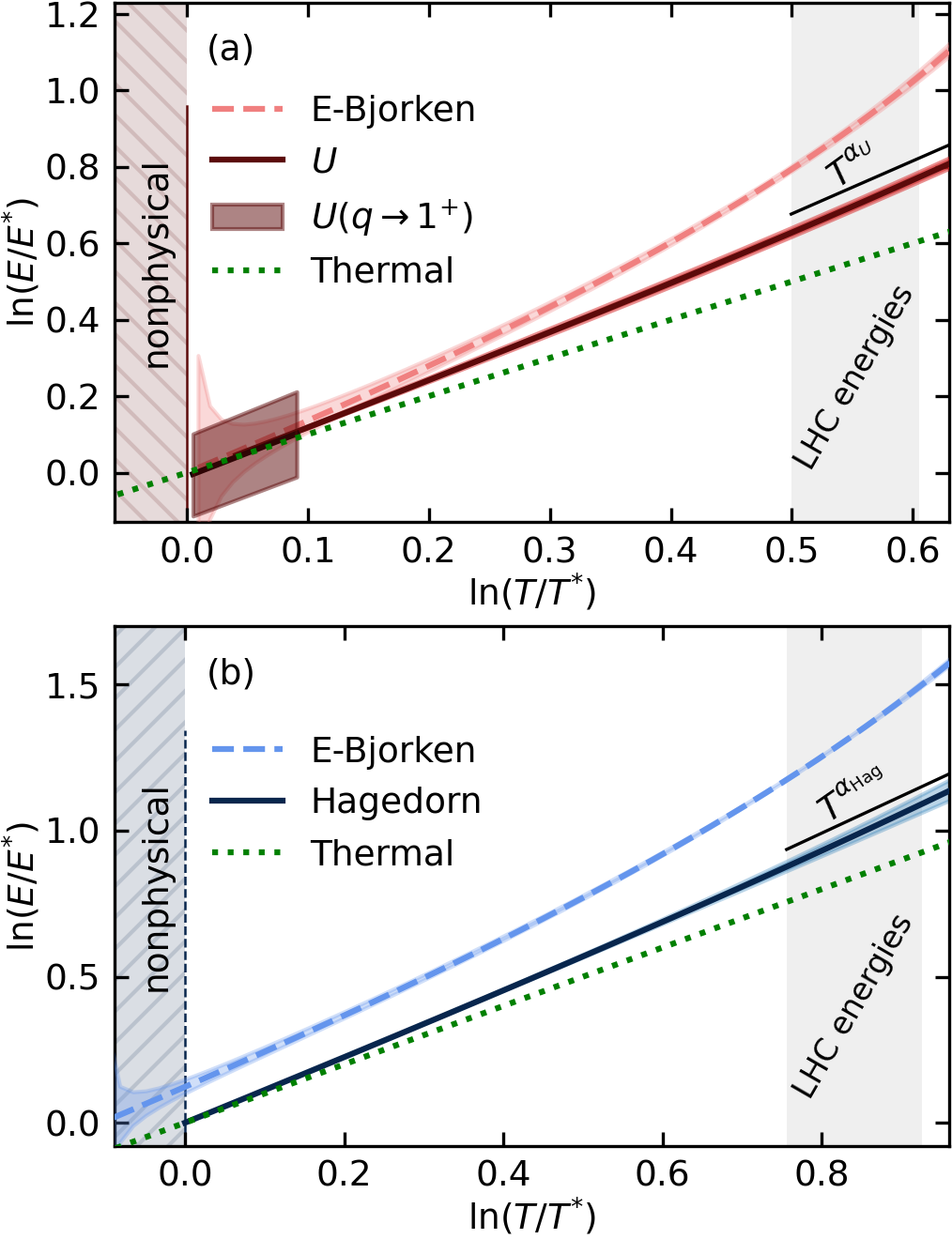}
    \caption{Internal energy $E$ scaled by $E^*$ as a function of $T/T^*$ for the (a) $U$ function and (b) Hagedorn function, plotted on logarithmic scales.
    Solid lines correspond to the estimation of the internal energy obtained from the Shannon entropy for the Hagedorn and $U$ functions.
    Dashed lines correspond to the Bjorken-type energy estimated as the average $\pT$.
In all cases, shaded regions correspond to the uncertainty propagation. 
The numerical calculations using the fourth-order Runge-Kutta method near the thermal limit ($q\to 1^{+}$) (bordered shaded regions) in the finite temperature interval correspond to the 95\% confidence level. 
    The dotted lines represent the thermal behavior.
    The hatched regions correspond to the nonphysical region for each model. }
    \label{fig:Energy}
\end{figure}

Note that the heat capacity is a positive increasing function, indicating that such systems are thermodynamically stable and that the number of degrees of freedom rises as the temperature increases.
Nevertheless, since our model does not explicitly incorporate the microscopic details of hadron composition, we can only speculate that the number of meson and baryon species increases as the bosonic degrees of freedom do. At higher energies, the heavy quark flavor production should contribute to raising the heat capacity and leading to the production of heavier hadrons.

The key role of heat capacity is that it gives us a fundamental relation to calculate the internal energy through the following differential equation:
\begin{equation}
    \frac{dE}{dT}=C,
\end{equation}
which is solved using the fourth-order Runge-Kutta method for each model. %\cite{Press:2007ipz}
For both the Hagedorn and Tricomi functions, we choose the starting integration point at $E^*=T^*$, which corresponds to the limit of an ideal gas.
Figure~\ref{fig:Energy} contains our results of the internal energy for the models discussed in this paper.
%For $T>T^*$, we found that the internal energy is a nonlinear function of the temperature, i.e., $E\propto T^\alpha$, with $\alpha_\text{Hag}=1.145(2)$ and $\alpha_U=1.266(2)$ for the Hagedorn and Tricomi functions, respectively.
We found that the internal energy follows a power law trend for temperatures in the range and above the LHC energies, i.e., $E\propto T^\alpha$, with $\alpha_\text{Hag}=1.267(4)$ and $\alpha_U=1.385(2)$ for the Hagedorn and Tricomi functions, respectively.
These results are consistent with the nonthermal and nonextensive description of the charged particle production in pp collisions.
We must emphasize that in all cases, the temperature of the system surpasses $T^*$, implying that the production of high $\pT$ hadrons is guaranteed, giving shape to the hard part of the $\pT$ spectrum.
In analogy with the thermodynamics of classical systems with vibrational modes, the total number of degrees of freedom (the equipartition theorem \cite{callen,handler1987introduction}) is not enough to establish a fundamental thermodynamic relation.

An alternative thermodynamics can be derived by setting the average $\pT$ as the energy per particle and retaining the temperature as the slope of the $\pT$ spectrum.
In this case, the goal is to determine a thermodynamically consistent entropy, which will be denoted as $S$ in order to avoid misleading with the Shannon entropy $H$.
This selection attempts to discuss an approach similar to the Bjorken energy density, without the problems encountered with the transverse area.
In this case, the thermal limit changes in the following way.
In a first instance, the internal energy takes the form $E=2T$, from which the heat capacity is $C=2$. 
This result is consistent with the discussion above.
In any case, the system requires more energetic collisions to produce particles with higher transverse momentum.
However, the doubling of the center of mass energy increment is required to double the average $\pT$, indicating that the system is described by twice the degrees of freedom compared to the thermodynamics derived from the Shannon entropy discussed above.
Regarding the entropy, it is found to be
\begin{equation}
    S=2\ln T +S_0,
\end{equation}
with $S_0$ being an arbitrary constant resulting from solving the second law of thermodynamics, and usually associated with the entropy of the ground state.

Now, let us discuss the nonthermal case.
We recall that the average $\pT$ for the Hagedorn and Tricomi models is given by Eqs.~\eqref{eq:avgpTHag} and \eqref{eq:avgpTU}, respectively. We plot this Bjorken-like energy in Fig.~\ref{fig:Energy} as dashed lines.
In contrast with the above estimation, the Bjorken-type energy can not be well described by a power law trend for the range temperature corresponding to the LHC energies.

Note that the average $\pT$ can be rewritten in a compact way as $\langle \pT \rangle=b(x)T$, where $b(x)$ represents the proportionality factor depending on the parameter $x=m, q$.
Therefore, the heat capacity is
\begin{equation}
    C=b(x(T))+T\frac{d b(x)}{dx}\frac{dx}{dT},
\end{equation}
where $x(T)$ denotes the dependence of $m$ and $q$ on $T$.
Figure~\ref{fig:ShannonHeat} contains our results on the heat capacity estimation for the Hagedorn and Tricomi functions.

In this case, the thermodynamically consistent entropy should be estimated by solving the following differential equation
\begin{equation}
    C=T\frac{dS}{dT},
    \label{eq:DES}
\end{equation}
with the initial condition $S^*=2\ln T^*+S_0$ and $E^* = 2T^*$ at $T^*$.
Similarly to the estimation of the energy in the case discussed above, Eq.~\eqref{eq:DES} is numerically solved using the fourth-order Runge-Kutta method.
Since $S_0$ is an arbitrary constant, it is fixed to zero. 
Moreover, to prevent misleading interpretations, we present our results of the entropy as the difference between the $S$ numerically obtained and the entropy for the ideal case, that is, we define $\Delta S=S-2\ln T -S_0$, canceling the contribution of $S_0$ and at the same time explicitly showing the functional behavior of the deviation of the entropy with respect to the ideal case.
In Fig.~\ref{fig:ShannonHeat}, we plot our estimations of $\Delta S$ as a function of $T$.
Similarly to $\Delta H$, $\Delta S$ is also a nonlinear increasing function.
Again, the main implication of the results obtained from the Bjorken-type energy is that the system experiences an increase in the degree of freedom that cannot be determined solely by applying the equipartition theorem.

%U: 1.38491 +/- 0.00163
%Hagedorn: 1.26650 +/- 0.00444

%E-Bjorken U: 2.36808 +/- 0.01961
%E-Bjorken Hagedorn: 1.95476 +/- 0.01116

%The internal energy behavior at the LHC energies and beyond in both approaches is described by $\propto T^{\alpha}$. 
%For the Shannon entropy way, we obtain $\alpha = 1.385(2)$ for the Tricomi function and $\alpha = 1.267(4)$ for the Hagedorn function.
%For the energy per particle inspired by the Bjorken energy: $\alpha = 2.37(2)$ and $\alpha = 1.95(1)$, for the Tricomi and Hagedorn functions, respectively.

\section{Conclusions}\label{sec:conclusions}

In this paper, we presented a thermostatistical analysis of the charged particle production in minimum bias pp collisions across LHC energies. 
To do this, we used the usual definition of the temperature, which is determined by analyzing the asymptotic behavior of the $\pT$ spectrum at low $\pT$ values. 
Additionally, the entropy was calculated using the Shannon entropy definition over the normalized $\pT$ spectrum, considering the entire $\pT$ range.
An alternative thermodynamic description can be derived by assuming the average $\pT$ as the energy estimator with the aim of discussing the Bjorken energy density approach.
In the analysis of the $\pT$ spectrum, we employed the Hagedorn and Tricomi functions.
Interestingly, these functions can be derived from the Schwinger mechanism by considering heavy tailed distributions of the string tension fluctuations, revealing that the medium where the charged particles are produced is no longer thermal.
A consequence of the nonextensivity of the charged particle production is nonthermal, with the temperature varying over small regions within the system. 
The main result reported in this paper is that the internal energy and entropy are nonlinear functions at LHC energies.
The main implication of this result is that the medium where the charged particles are formed deviates from the thermal description, and the count of degrees of freedom is more complex than simply adding the number of flavors producing heavier particles.
%This affirmation is supported by our results on the internal energy estimation, which behaves as a nonlinear function of temperature ($E\propto T^\alpha$) at LHC energies and becomes linear ($E=T$) in the thermal limit, which resembles the ideal gas model and suggests the existence of a thermal-nonthermal transition temperature for both approaches, the Hagedorn and Tricomi functions. 
It is remarkable that, in this approach, the internal energy is determined from the experimental data through thermodynamic relations, rather than requiring detailed microscopic assumptions about the QCD interactions.
%Our estimation of $T_\text{Hag}^*=0.63 T_U^*$ and $T_U^*=0.103(6)$ GeV occurs at the center of mass energy slightly greater than $2 m_p$.
This result marks the onset of nonextensive behavior in the $\pT$ spectrum having a power-law tail, characterizing the hard processes involved in the production of high $\pT$ particles. 
This explains why the hard part of the $\pT$ spectrum (production of high $\pT$ hadrons) has been observed experimentally in the production of charged particles in pp collisions at very low center of mass energies and may also encompass other particle collisions such as e$^-$p or e$^+$e$^-$, for which the production of high $\pT$ hadrons has also been reported.

%\bmhead{Supplementary information}

%If your article has accompanying supplementary file/s please state so here. 

%Authors reporting data from electrophoretic gels and blots should supply the full unprocessed scans for key as part of their Supplementary information. This may be requested by the editorial team/s if it is missing.

%Please refer to Journal-level guidance for any specific requirements.

\acknowledgments
This work was funded by Secretaría de Ciencia, Humanidades, Tecnología e Innovación (SECIHTI-México) under the project CF-2019/2042, graduate fellowship grant number 1140160, and postdoctoral fellowship grant numbers 645654 and 289198.
C.P. is supported by Xunta de Galicia (CIGUS Network of Research Centers), Spanish Research Agency under the project PID2023-152762NB, and ``Maria de Maeztu'' Units of Excellence program CEX2023-001318-M.

\bibliography{ref}% common bib file

\end{document}